\documentclass[letterpaper, 10pt, conference]{ieeeconf}  

\IEEEoverridecommandlockouts                              

\usepackage{graphicx}
\usepackage{url}  
\usepackage{amsmath}
\usepackage{todonotes}

\begin{document}

\title{\LARGE \bf
	Low-complexity deep learning frameworks for acoustic scene classification
}

\author{Technical Report for DCASE 2022 Task 1 challenge (AIT-Essex-Teams)\\
Lam Pham, Dat Ngo, Anahid Jalali, Alexander Schindler
}

\maketitle
\thispagestyle{empty}
\pagestyle{empty}

\begin{abstract}
In this report, we presents low-complexity deep learning frameworks for acoustic scene classification (ASC). 
The proposed frameworks can be separated into four main steps: Front-end spectrogram extraction, online data augmentation, back-end classification, and late fusion of predicted probabilities. 
In particular, we initially transform audio recordings into Mel, Gammatone, and CQT spectrograms.
Next, data augmentation methods of Random Cropping, Specaugment, and Mixup are then applied to generate augmented spectrograms before being fed into deep learning based classifiers.
Finally, to achieve the best performance, we fuse probabilities which obtained from three individual classifiers, which are independently-trained with three type of spectrograms.
Our experiments conducted on DCASE 2022 Task 1 Development dataset have fullfiled the requirement of low-complexity and achieved the best classification accuracy of 60.1\%, improving DCASE baseline by 17.2\%.

\indent \textit{Clinical relevance}--- Mixup data augmentation, Convolutional Neural Network (CNN), pruning, quantization, spectrogram, Gammatone filter.

\end{abstract}

\section{Introduction}
\label{intro}

To deal with one of the main ASC challenges, mismatched recording devices, a variety of methods have been proposed, which mainly make use of ensemble techniques: Ensemble of spectrogram inputs~\cite{lam01, lam03, phan2019spatio, lam04, lam05, truc_dca_18, yuma, huy_mul, lam02, pham2022wider, pham2021deep} (i.e., This approach uses multiple spectrogram inputs but only one model architecture) or ensemble of different classification models~\cite{phaye_dca_18, zhao_dca_17, hong_dca_18} (i.e., This approach uses only one spectrogram input, but explores the spectrogram by different model architectures).
Although these approaches help to achieve good results, they show very large footprint models, which causes challenging to apply on edge-devices.
Indeed, all top-10 systems proposed in recent DCASE challenges in  2018, 2019, 2020 present large architectures of deep neural networks, requiring larger than 2 MB of memory to store trainable parameters.
Recently, DCASE 2021 and DCASE 2022 Task 1A challenges~\cite{task1a_2022} focus on dealing the issue of high-complexity model, then require the maximum model complexity of 128 KB.
Notably, DCASE 2022 Task 1A challenge does not allow to use pruning techniques as the pruning parameters still occupy the memory and cost the computation on edge-devices.

In this report, we introduces robust and low-complexity deep learning frameworks for ASC task.
In particular, to deal with the ASC challenge of mismatched recording devices, we propose an ensemble of multiple spectrogram inputs, using Mel filter~\cite{librosa_tool}, Gammatone~\cite{aud_tool} filter, and CQT~\cite{librosa_tool}.
For each network used for training an individual spectrogram input, we deal with the issue of model complexity by applying multiple techniques of channel reduction, decomposed convolution, and quantization.  

\section{The proposed low-complexity deep learning framework}
\begin{figure}[th]
	\centering
    \scalebox{0.95}{
	\centerline{\includegraphics[width=\linewidth]{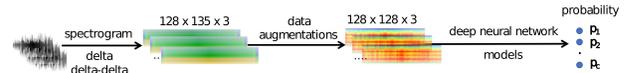}}
	}
	\vspace{0.1cm}
	\caption{High-level architecture of an ASC system.}
	\label{fig:high_level_arc}
\end{figure}

\subsection{Propose low-complexity ASC models}
A framework for ASC task with high-level architecture is shown in Fig.~\ref{fig:high_level_arc}.
Initially, a draw audio signal is firstly transformed into a spectrogram of  128$\times$135 by using MEL filter~\cite{librosa_tool}, Gammatone filter~\cite{aud_tool}, or CQT~\cite{librosa_tool} with the FFT number, Hanna window size, hop size, and the filter number set to 4096, 2048, 306, and 128.
We then apply delta and delta-delta on the spectrogram, generate a new spectrogram of 128$\times$135$\times$3 (i.e., The channel dimension is 3 which causes by concatenating the original spectrogram, delta, and delta-delta).

Next, we apply data augmentation methods on the spectrograms.
In this report, we apply three data augmentation methods of Random Cropping~\cite{aug_crop}, Specaugment~\cite{aug_spec}, and Mixup~\cite{mixup1, mixup2}. 
In particular, the temporal dimension of spectrograms of 128$\times$\textbf{135}$\times$3 is randomly cropped into 128$\times$\textbf{128}$\times$3 (e.g., Random Cropping method). 
Then, ten continuous and random frequency or temporal bins of the cropped spectrograms are erased (e.g., Specaugment method). 
Finally, the spectrograms are randomly mixed together using different ratios from Uniform or Beta distributions (e.g., Mixup method). 
All of three data augmentation methods are applied on each batch of spectrograms during the training process, referred to as the online data augmentation methods.

We then fed augmented spectrograms into a back-end deep learning networks for classification.
We propose three models in this report: (M1) a  model with four convolutional layers as shown in Table~\ref{table:M1}, (M2)  a model with six convolutional layers as shown in Table~\ref{table:M2}, and (M3) a model with eight convolutional layers as shown in Table~\ref{table:M3}.
As using limited filter numbers, proposed models present low complexity architectures, in which M3 presents the largest one.

\begin{figure}[t]
	\centering
	\centerline{\includegraphics[width=\linewidth]{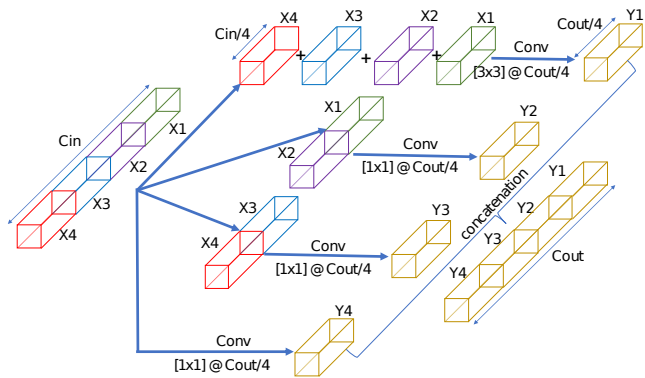}}
	\vspace{0.1cm}
	\caption{Decomposed convolution technique applied to a convolutional layer.}
	\label{fig:decompose}
\end{figure}

Inspired by~\cite{pham2021low}, the convolutional layer used in these models is decomposed to sub convolutional computation as shown in Figure~\ref{fig:decompose}. 
By decomposing a convolutional layer into four sub-convolutional layers, the model complexity is reduced to nearly 1/8.5.
Additionally, 8-bit quantization method is applied on each model which further reduces the model size to 1/4.
As a result, three proposed low-footprint models report the model complexity of 10.6 KB, 10.0 KB, and 36.8 KB for M1, M2, and M3, respectively.

\begin{table}[t]
    \caption{Low-complexity M2} 
        	\vspace{-0.2cm}
    \centering
    \scalebox{0.7}{
    \begin{tabular}{|l |c|} 
        \hline 
            \textbf{Network architecture}   &  \textbf{Output}  \\
        \hline 
         Convolution ([$3{\times}3$] $@$  C\_{out}1=16) -  ReLU - BN - Dropout (10\%) & $128{\times}128{\times}16$\\
         
         Convolution ([$3{\times}3$] $@$  C\_{out}2=16) - ReLU - BN - AP [$2{\times}2$] - Dropout (10\%) & $64{\times}64{\times}16$\\
         
         Convolution ([$3{\times}3$] $@$ C\_{out}3=32) - ReLU - BN - Dropout (10\%) & $64{\times}64{\times}32$\\
         
         Convolution ([$3{\times}3$] $@$ C\_{out}4=32) - ReLU - BN - AP [$2{\times}2$] - Dropout (10\%) & $32{\times}32{\times}32$\\
         
         Convolution ([$3{\times}3$] $@$ C\_{out}5=64) - ReLU - BN - AP [$2{\times}2$] - Dropout (10\%) & $16{\times}16{\times}64$\\
         
         Convolution ([$3{\times}3$] $@$ C\_{out}6=64) - ReLU - BN - GAP - Dropout (10\%) & 64\\
                  
         FC - Softmax  &  $C=10$       \\
       \hline 
    \end{tabular}
    }
    \label{table:M2} 
\end{table}
\begin{table}[t]
    \caption{Low-complexity M3} 
        	\vspace{-0.2cm}
    \centering
    \scalebox{0.7}{
    \begin{tabular}{|l |c|} 
        \hline 
            \textbf{Network architecture}   &  \textbf{Output}  \\
        \hline 
         Convolution ([$3{\times}3$] $@$  C\_{out}1=16) -  ReLU - BN - Dropout (10\%) & $128{\times}128{\times}16$\\
         
         Convolution ([$3{\times}3$] $@$  C\_{out}2=16) - ReLU - BN - AP [$2{\times}2$] - Dropout (10\%) & $64{\times}64{\times}16$\\
         
         Convolution ([$3{\times}3$] $@$ C\_{out}3=32) - ReLU - BN - Dropout (10\%) & $64{\times}64{\times}32$\\
         
         Convolution ([$3{\times}3$] $@$ C\_{out}4=32) - ReLU - BN - AP [$2{\times}2$] - Dropout (10\%) & $32{\times}32{\times}32$\\
         
         Convolution ([$3{\times}3$] $@$ C\_{out}5=64) - ReLU - BN - AP [$2{\times}2$] - Dropout (10\%) & $32{\times}32{\times}64$\\
         
         Convolution ([$3{\times}3$] $@$ C\_{out}6=64) - ReLU - BN - AP - Dropout (10\%) & $16{\times}16{\times}64$\\
        
        Convolution ([$3{\times}3$] $@$ C\_{out}6=128) - ReLU - BN - AP - Dropout (10\%) & $16{\times}16{\times}128$\\

        Convolution ([$3{\times}3$] $@$ C\_{out}6=128) - ReLU - BN - GAP - Dropout (10\%) & 128\\
                  
         FC - Softmax  &  $C=10$       \\
       \hline 
    \end{tabular}
    }
    \label{table:M3} 
\end{table}
\begin{table}[t]
    \caption{Low-complexity M1} 
        	\vspace{-0.2cm}
    \centering
    \scalebox{0.7}{
    \begin{tabular}{|l |c|} 
        \hline 
            \textbf{Network architecture}   &  \textbf{Output}  \\
        \hline 
         Convolution ([$3{\times}3$] $@$  C\_{out}2=16) - ReLU - BN - AP [$2{\times}2$] - Dropout (10\%) & $64{\times}64{\times}16$\\
         
         Convolution ([$3{\times}3$] $@$ C\_{out}4=32) - ReLU - BN - AP [$2{\times}2$] - Dropout (10\%) & $32{\times}32{\times}32$\\
         
         Convolution ([$3{\times}3$] $@$ C\_{out}5=64) - ReLU - BN - AP [$2{\times}2$] - Dropout (10\%) & $16{\times}16{\times}64$\\
         
         Convolution ([$3{\times}3$] $@$ C\_{out}6=128) - ReLU - BN - GAP - Dropout (10\%) & 128\\
                  
         FC - Softmax  &  $C=10$       \\
       \hline 
    \end{tabular}
    }
    \label{table:M1} 
\end{table}

As using multiple spectrograms as a rule of thumb to improve ASC performance~\cite{lam01, lam02, lam03, lam04}, an ensemble of three spectrograms of log-Mel, Gammatone, and CQT is conducted in this work. 
As we need to train independently three models for three different spectrogram inputs, the final complexity of the proposed frameworks (i.e., each framework comprises three models of M1, M2 or M3) are approximately 31.8 KB, 30 KB, and 110 KB respectively, which meets DCASE 2022 Task 1 challenge requirement (i.e., the final model must be less than 128 KB).

To fuse probability results obtained from three spectrograms, we conduct experiments over individual networks with each spectrogram input, then obtain predicted probability of each network as  \(\mathbf{\bar{p_{s}}}= (\bar{p}_{s1}, \bar{p}_{s2}, ..., \bar{p}_{sC})\), where $C$ is the category number and the \(s^{th}\) out of \(S\) networks evaluated. 
Next, the predicted probability after PROD fusion \(\mathbf{p_{f-prod}} = (\bar{p}_{1}, \bar{p}_{2}, ..., \bar{p}_{C}) \) is obtained by:

\begin{equation}
\label{eq:mix_up_x1}
\bar{p_{c}} = \frac{1}{S} \prod_{s=1}^{S} \bar{p}_{sc} ~~~  for  ~~ 1 \leq s \leq S 
\end{equation}

Finally, the predicted label  \(\hat{y}\) is determined by \begin{equation}
    \label{eq:label_determine}
    \hat{y} = arg max (\bar{p}_{1}, \bar{p}_{2}, ...,\bar{p}_{C} )
\end{equation}

\section{Evaluation Setting and Results}

\subsection{TAU Urban Acoustic Scenes 2022 Mobile, development dataset~\cite{dc_2021_1A}}
This dataset is referred to as DCASE 2022 Task 1 Development, which was proposed for DCASE 2022 challenge~\cite{dcase_web}. 
In this challenge, the limitation of model complexity is set to 128 KB of trainable parameter, not allow to use pruning techniques, and evaluate on 1-second audio segment.
The dataset is slightly unbalanced, being recorded across 12 large European cities: Amsterdam, Barcelona, Helsinki, Lisbon, London, Lyon, Madrid, Milan, Prague, Paris, Stockholm, and Vienna. 
It consists of 10 scene classes: airport, shopping mall (indoor), metro station (underground), pedestrian street, public square, street (traffic), traveling by tram, bus and metro (underground), and urban park.
The audio recordings were recorded from 3 different physical devices namely A (10215 recordings), B (749 recordings), C (748 recordings).
Additionally, synthetic data for mobile devices was created based on the original recordings, referred to as S1 (750 recordings), S2 (750 recordings), S3 (750 recordings), S4 (750 recordings), S5 (750 recordings), and S6 (750 recordings). 
  
To evaluate, we follow the DCASE 2022 Task 1 challenge~\cite{dcase_web}, use two sub-sets known as Training (Train.) and Evaluation (Eval.) from the Development set for training and testing processes, respectively.
Notably, two of 12 cities and S4, S5, S6 audio recordings are only presented in the Eval. subset for evaluating the issue of mismatched recording devices and unseen samples. 

\subsection{Deep learning framework implementation}
We use Tensorflow framework to construct all deep learning models proposed in this report and Adam method~\cite{adam} for optimization.
The training and evaluating processes are conducted on GPU Titan RTX 24GB.
As we use Mixup for data augmentation method, labels are not one-hot encoding format. Therefore, Kullback–Leibler divergence (KL) loss~\cite{kl_loss} is used as shown in Eq. (\ref{eq:kl_loss}) below.
\begin{align}
   \label{eq:kl_loss}
   Loss_{KL}(\Theta) = \sum_{n=1}^{N}\mathbf{y}_{n}\log \left\{ \frac{\mathbf{y}_{n}}{\mathbf{\hat{y}}_{n}} \right\}  +  \frac{\lambda}{2}||\Theta||_{2}^{2}
\end{align}
where  \(\Theta\) are trainable parameters, constant \(\lambda\) is set initially to $0.0001$, $N$ is batch size set to 100, $\mathbf{y_{i}}$ and $\mathbf{\hat{y}_{i}}$  denote expected and predicted results.

\subsection{Performance comparison between DCASE baseline and the proposed models}
As Table~\ref{table:bs_cmp} shows, all proposed ensembles in three models (M1, M2 and M3) outperform DCASE baseline across individual categories.
Ensemble at M3 model achieves the best overall performance of 60.1\% which improves DCASE baseline by 17.2\%.
Notably, `street\_pedestrian' class shows low performance compared with the others, which need to further investigate.

\section{Conclusion}

We have just presented low-complexity frameworks for ASC task, which makes use multiple spectrogram inputs and model compression techniques.
While the ensemble of multiple spectrograms helps to tackle different ASC challenges of mismatched recording devices or lacking of input to improve the performance, multiple techniques of model compression help to achieve low-complexity models less than 128 KB. 

\begin{table}[t]
    \caption{Performance comparison among DCASE baseline, the proposed ensemble of multiple spectrograms with models M1, M2, or M3} 
        	\vspace{-0.2cm}
    \centering
    \scalebox{1.0}{
    \begin{tabular}{| l |  c c c c  | } 
        \hline 
          \textbf{Category}       &\textbf{DCASE}     &\textbf{Ensemble}    &\textbf{Ensemble}  &\textbf{Ensemble}  \\
          \textbf{}       &\textbf{baseline}     &\textbf{at M1}    &\textbf{at M2}  &\textbf{at M3}  \\
        \hline 

	    Airport           &39.4           &52.5           &60.7  &57.2          \\
        Bus               &29.3           &69.4           &62.5  &71.3          \\
        Metro             &47.9           &44.9           &36.2  &44.9        \\
        Metro station     &36.0           &50.6           &28.3  &40.9       \\
        Park              &58.9           &81.1           &67.2  &83.9        \\
        Public square     &20.8           &23.8           &37.7  &50.8         \\
        Shopping mall     &51.4           &74.2           &60.7  &60.5         \\
        Street pedestrian &30.1           &27.9           &23.0  &34.1         \\
        Street traffic    &70.6           &81.2           &78.6  &78.8         \\
        Tram              &44.6           &50.5           &59.1  &78.0         \\
        \hline                                                       
        Average           &42.9           &55.6           &51.4  &60.1         \\
        \hline                                                       

    \end{tabular}
    }
    \label{table:bs_cmp} 
\end{table}
%

\bibliographystyle{IEEEbib}
\bibliography{refs}


\end{document}